\documentclass[twocolumn,preprintnumbers,amsmath,amssymb,superscriptaddress]{revtex4-2}
\UseRawInputEncoding

\usepackage{latexsym,amssymb,amsthm,amsmath,epsfig,braket}
\usepackage[left=2.00cm, right=2.00cm, top=2.00cm, bottom=2.00cm]{geometry}
\usepackage{xcolor}
\usepackage{svg}
\usepackage[colorlinks, citecolor={blue!80!black}, urlcolor={blue!80!black}, linkcolor={red!50!black}]{hyperref}
\usepackage{float}
\usepackage{hyperref}
\hypersetup{
    citecolor=red,
    colorlinks=true,
    linkcolor=blue,
    filecolor=blue,      
    urlcolor=blue,
}

\begin{document}

\title{\textbf{Coherent control of photonic spin Hall effect in a Cavity}}

\author{Muzamil Shah}
\affiliation{Department of Physics, Zhejiang Normal University, Jinhua, Zhejiang 321004, China}
\affiliation{Department of Physics, Quaid-i-Azam University, Islamabad 45320, Pakistan}
\author{Shahid Qamar}
\affiliation{Department of Physics and Applied Mathematics, Pakistan Institute of Engineering and Applied Sciences (PIEAS), Nilore $45650$, Islamabad, Pakistan}
\affiliation{Center for Mathematical Sciences, PIEAS, Nilore, Islamabad $45650$, Pakistan}
\author{Muhammad Waseem}
\email{mwaseem@pieas.edu.pk}
\affiliation{Department of Physics and Applied Mathematics, Pakistan Institute of Engineering and Applied Sciences (PIEAS), Nilore $45650$, Islamabad, Pakistan}
\affiliation{Center for Mathematical Sciences, PIEAS, Nilore, Islamabad $45650$, Pakistan}

\date{\today}

% use {abstract*} to suppress the copyright line. Copyright information will be added in production

\begin{abstract} 
This paper theoretically investigates the manipulation of the Photonic Spin Hall Effect (photonic SHE) using a four-level closed
coherent control coupling scheme in a cavity.
The atomic system is configured to function as a combined Tripod and Lambda (CTL), Lambda $\Lambda$, and $N$ level model by properly adjusting the control field strengths and their relative phases.
The system demonstrates multiple transparency windows in the CTL configuration, allowing the tunable photonic SHE to be used over a wider range of probe field detuning. 
At probe field resonance under the condition of electromagnetic induced transparency (EIT), the $\Lambda$-type system exhibits photonic SHE similar to the CTL system, showing a maximum upper limit of photonic SHE equal to half of the incident beam waist. This upper limit arises due to zero absorption and dispersion. Control field strengths and atomic density do not influence photonic SHE at resonance for both atomic configurations.
Our findings reveal that atomic density and strength of control fields significantly influence photonic SHE in the $N$-type model at resonance, offering additional control parameters for tuning photonic SHE. 
Finally, the results are equally valid and applicable to conventional $\Lambda$-type and N-type atomic systems, making the findings broadly relevant in cavity atomic systems. The results of angular photonic SHE are also discussed.

\end{abstract}
\maketitle
%%%%%%%%%%%%%%%%%%%%%%%%%%  body  %%%%%%%%%%%%%%%%%%%%%%%%%%

\section{Introduction}
Originally, the Spin Hall Effect (SHE) arises in solid-state systems, where it refers to the transverse shift of particles (such as electrons) due to spin-orbit coupling \cite{kato2004observation, sinova2015spin, bernevig2006quantum}. 
In these electronic systems, the spin-orbit interaction causes particles with spin-up and spin-down states to experience different transverse forces when moving through the electronic potential \cite{jungwirth2012spin}. As a result, the particles displace in the transverse direction known as the SHE \cite{hirsch1999spin}. 
The SHE has been widely studied in condensed matter systems such as semiconductors \cite{kato2004observation}, graphene \cite{kane2005quantum},  topological insulators \cite{bernevig2006quantum}, and two-dimensional materials \cite{safeer2019room}. In these condensed matter systems, spin separation occurs due to intrinsic or induced mechanisms.
Similarly, spin separation has been achieved through optically generated spin currents of exciton-polaritons in semiconductor microcavities, a phenomenon known as optical SHE \cite{kavokin2005optical}.
\\
On the other hand, the photonic SHE deals with the analogous behavior of photons in coherent light-matter interaction.
In photonic SHE, right-circularly polarized and left-circularly polarized components play a role similar to spin-up and spin-down electrons. 
The refractive index gradient of the matter acts as an analog to the electronic potential.
In photonic SHE, left-circularly polarized and right-circularly polarized photons experience different shifts perpendicular to the incident plane due to spin-orbit interaction as they interact with the interface of the coherent medium \cite{kim_spin_2023}.
The photonic SHE is mainly attributed to the optical angular momentum and two geometric phases \cite{sheng2023photonic}. 
One geometric phase is the spin-redirection Rytov-Vlasimirskii-Berry (RVB) phase associated with the propagation direction of the wave vector, and the second is the Pancharatnam-Berry (PB) phase related to the polarization manipulation of light \cite{sheng2023photonic,liu2022photonic}. 
Recently, the photonic SHE gained particular attention for its ability to control spin-dependent behaviors of photons in various optical media. For example, semiconductors~\cite{menard_imaging_2009}, graphene layers~\cite{zhou_identifying_2012,PhysRevA.95.013809,SHAH2024107676}, surface plasmon resonance systems~\cite{salasnich_enhancement_2012,zhou_enhanced_2016,tan_enhancing_2016,xiang_enhanced_2017,wan_controlling_2020}, metamaterials~\cite{yin_photonic_2013}, all-dielectric metasurfaces~\cite{kim_reaching_2022}, topological insulators \cite{SHAH2022115113}, strained Weyl semimetals \cite{jia2021tunable}, hyperbolic metamaterials \cite{kapitanova2014photonic}, atomic medium \cite{Aizaz2025,Munir2025} and two-dimensional quantum materials \cite{Shah_2022,kort-kamp_topological_2017}.
The photonic SHE has enabled applications such as probing topological phase transitions~\cite{Shah_2022, kort-kamp_topological_2017}, chiral molecular detection~\cite{tang_optimal_2023}, spin switches \cite{sheng2025effectively}, performing mathematical operations and edge detection~\cite{zhu_generalized_2019}. The spin-dependent shift has been extensively
investigated for applications in precision metrology \cite{zhou_identifying_2012,PhysRevA.85.043809,chen2020precision,wang2020ultrasensitive} and differential phase imaging~\cite{tang2025observation,zhao2025optical,yang2025quantum}. Furthermore, the spatial shifts can sensitively probe physical parameters, such as measuring the surface susceptibility of single-layer crystals \cite{chen2021measurement} and the optical constants of monolayer $\mathrm{MoS_{2}}$ via the photonic spin Hall effect \cite{zheng2024measurement}.

The photonic SHE traces back to the out-of-plane Imbert-Fedorov (IF) shift perpendicular to the incident plane \cite{imbert_calculation_1972}.
The IF Shift can also be attributed to the spin-orbit interaction of light and is linked with the spin-redirection RVB phase \cite{sheng2023photonic,liu2022photonic}. However, IF-Shift is formulated differently by considering right-circular or left-circular polarized light at the optical interface~\cite{liu2022photonic}. On the other hand, photonic SHE considers the incident linearly polarized beam as a superposition of the right-circular polarized and left-circular polarized components.
Besides, an in-plane Goos-Hanchen (GH) shift to the incident plane at an optical interface originates from the spatial dispersion of beam reflection, transmission coefficients, and the interference of the angular spectrum components~\cite{goos1947neuer}.
\\
In quantum optics, atomic coherence effects of light-matter interaction have enabled groundbreaking discoveries, including electromagnetically induced transparency (EIT) \cite{fleischhauer2005electromagnetically,boller_observation_1991}, lasing without inversion \cite{scully1989degenerate}, low and ultraslow light propagation \cite{hau1999light,budker1999nonlinear}, stationary light \cite{bajcsy2003stationary}, light storage \cite{phillips2001storage,liu2001observation}, amplified nonlinear optical effects \cite{schmidt1996giant,pack2006transients}, GH shift \cite{ziauddin2010coherent}, and so on.
In most of these studies, the coherent atomic medium is considered a three-level $\Lambda$, $V$, ladder, and a four-level $N$-type energy level structure. 
Recently, a novel atom-light coupling scheme, known as the combined tripod and $\Lambda$ (CTL) scheme, was introduced for electromagnetically induced transparency (EIT) and slow light \cite{hamedi2017electromagnetically}.
Changing the amplitudes and phases of the control fields transforms the CTL system into both $\Lambda$ and $N$-type systems. 
This closed-loop control field structure makes the CTL system a versatile model for achieving multiple interaction pathways and quantum interference effects (CTL, $\Lambda$, and $N$-type) within the same atomic ensemble. 
Recent studies explored EIT \cite{hamedi2017electromagnetically}, structured light detection \cite{hamedi2018azimuthal}, and GH shift \cite{abbas2024goos} in the CTL atomic system. 
\\

Atomic coherence offers a possibility for controlling photonic SHE through coherent light-matter interactions. In previous work, we explored tunability of the photonic SHE in a $\Lambda$-type atomic medium using a gain-assisted mechanism, where the atom-field interaction is mediated by a two-photon Raman transition~\cite{waseem2024}.
Recent studies have shown that a four-level tripod atom-light coupling scheme enables dual-channel enhancement of the photonic SHE for both probe and signal fields under resonance conditions~\cite{muqadar2025, ABBAS2025131930}. Additionally, photonic SHE tunability has been explored in a system of three $\Lambda$-type atomic media embedded in an optomechanical cavity~\cite{akhter}, revealing that photon-phonon interactions allow dynamic control of the photonic SHE. 
In this paper, we investigate the tunability of the photonic SHE using a CTL atomic system within the cavity.
The CTL system integrates $\Lambda$-type EIT and $N$-type systems into a unified framework, providing a comprehensive approach to understanding their behavior. By adjusting the control field strengths and the relative phase, the CTL-atomic system can be reconfigured to $\Lambda$-type or an N-type atomic model.
The CTL configuration exhibits multiple transparency windows, providing an opportunity to achieve a tunable photonic SHE over a wider range of probe field detuning.  
In the $\Lambda$-type system, the photonic SHE exhibits behavior resembling the CTL configuration at probe field resonance.
The CTL and $\Lambda$ systems exhibit maximum photonic SHE due to zero absorption and dispersion at probe field resonance.   
Furthermore, atomic density and amplitude of control fields do not influence photonic SHE at probe field resonance. 
In the N-type system, the atomic density and the control field strength play a significant role in modulating the photonic SHE at probe field resonance.
At lower atomic densities, the photonic SHE \textbf{increases}, unlike in the CTL and $\Lambda$ configurations, where atomic density has no impact on the photonic SHE. 
Importantly, the results obtained from this study are equally valid and applicable to conventional $\Lambda$-type and N-type atomic systems, making the findings broadly relevant in cavity atomic systems.  

The rest of the paper is organized as follows: Section II presents a detailed theoretical model, and Section III discusses the results and analysis of the CTL, $\Lambda$-type, and N-type configurations.  
Finally, the conclusions are summarized in Section IV.

%%%%%%%%%%%%%%%%%%%%%%%%%%%%%%%%%%%%%%%%%%%%%%%%%%%%%%%%%%%%%%%%%%%%%%%%%%%%%
%%%%%%%%%%%%%%%%%%%%%%%%%%%%%%%%%%%%%%%%%%%%%%%%%%%
\begin{figure*}[t]
    \centering
    \includegraphics[width=0.85\linewidth]{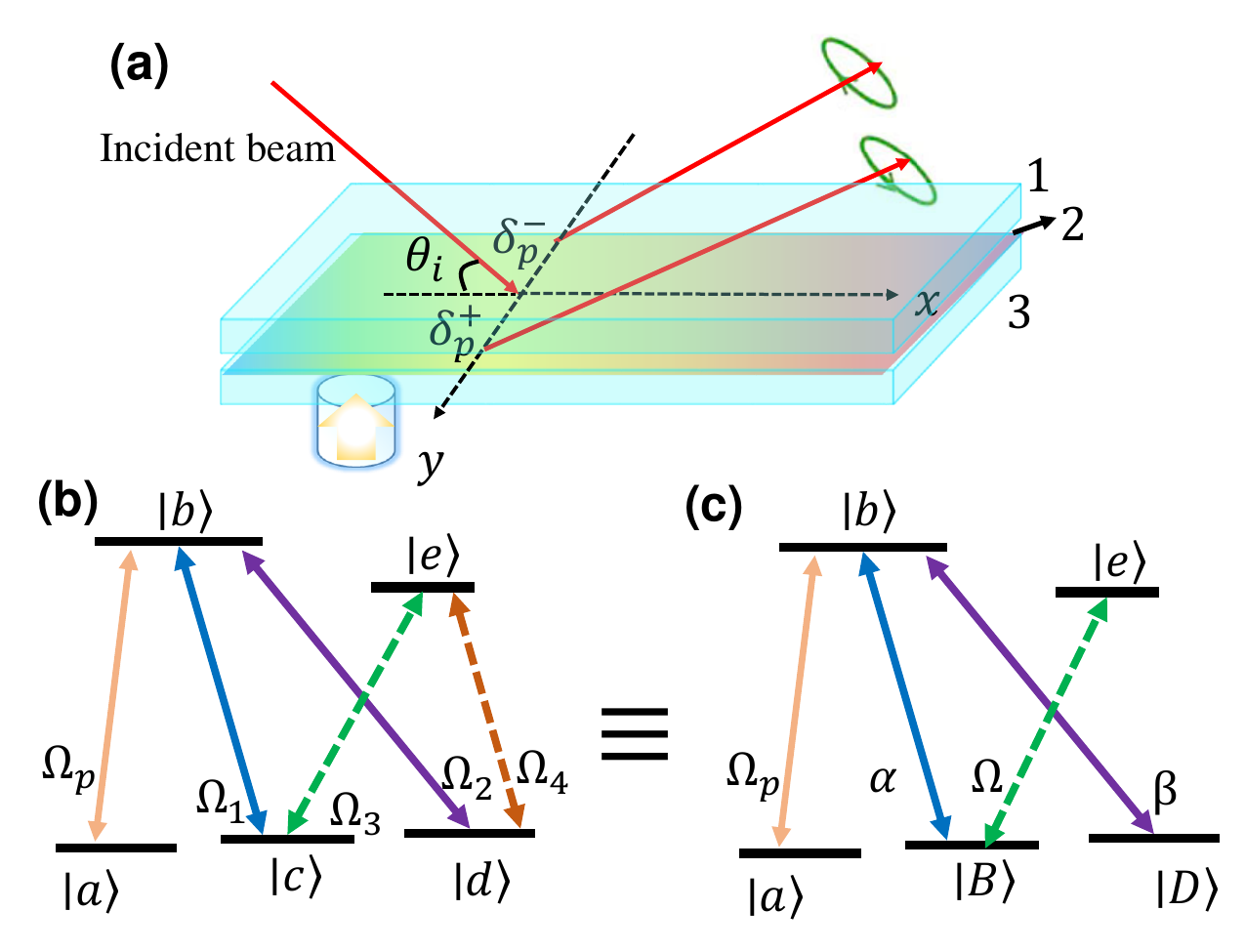}
    \caption{(a) Schematic diagram of the physical model. The atomic sample from the bottom nozzle spreads into an ultra-high vacuum glass vapor cell or cavity made of Pyrex with inner thickness $d=0.4~\mu$m. The TE and TM-polarized probe light beam is incident on the upper layer of the glass cell at an incident angle $\theta_{i}$. $\delta_p^{ +}$ and $\delta_p^{-}$ indicate the transverse displacements for right- and left-circular polarization components, respectively. (b) Energy level diagram of the five-level atomic system that acts as a combined Tripod and $\Lambda$ system known as the CTL-model. (c) The equivalent four-level diagram using internal dark and bright states.}
\label{fig:sys}
\end{figure*}
%%%%%%%%%%%%%%%%%%%%%%%%%%%%%%%%%%%%%%%%%%%%%%%%%%%%%%
%%%%%%%%%%%%%%%%%%%%%%%%%%%%%%%%%%%%%%%%%%%%%%%%

\section{\label{sec:level2} Atomic Model and Equations} 
We consider a five-level atomic system to obtain the dielectric susceptibility of the intra-cavity medium.
The energy level diagram of the five-level atomic system is shown in Fig.~\ref{fig:sys}(b), which consists of three ground states, $\ket{a}$, $\ket{c}$, $\ket{d}$, and two excited states $\ket{b}$ and $\ket{e}$. This five-level system combines tripod and $\Lambda$ (CTL) subsystems.  
The Tripod subsystem consists of three ground states, $\ket{a}$, $\ket{c}$, $\ket{d}$, and an excited state $\ket{b}$, while the three-level $\Lambda$ subsystem consists of two ground states, $\ket{c}$ and $\ket{d}$, and an excited state $\ket{e}$.
The probe field of the Rabi frequency $\Omega_p$ is applied to the transition $\ket{a} \longrightarrow \ket{b}$, the four control fields of Rabi frequencies $\Omega_{1}$, $\Omega_{2}$, $\Omega_{3}$, and $\Omega_{4}$ established the connection between transition $\ket{b} \longrightarrow \ket{c}$, $\ket{b} \longrightarrow \ket{d}$, $\ket{e} \longrightarrow \ket{c}$, and $\ket{e} \longrightarrow \ket{d}$, respectively. 
The Rabi frequencies of the control fields are complex and defined as $\Omega_j = |\Omega_j| e^{i \phi_j}$ with $j=1,2,3,4$. Here, $|\Omega_j|$ is the amplitude and $\phi_j$ is the phase of the $j^{th}$ control field.
Two distinct pathways $\ket{b} \longrightarrow \ket{c} \longrightarrow \ket{e}$ and $\ket{b} \longrightarrow \ket{d} \longrightarrow \ket{e}$ connect the two excited states $\ket{b}$ and $\ket{e}$. This connection creates a closed four-level coherent control coupling scheme by defining the relative phase $\phi=(\phi_1 - \phi_2) - (\phi_3 - \phi_4)$. The total interaction picture Hamiltonian of the system becomes  
\begin{align}
 H=&-\Omega^*_{p}\ket{a}\bra{b}-\Omega^*_{1}\ket{c}\bra{b}-\Omega^*_{2}\ket{d}\bra{b} \nonumber\\
 &-\Omega^*_{3}\ket{c}\bra{e}-\Omega^*_{4}\ket{d}\bra{e}- H.c.   
\end{align}
The two ground states $\ket{c}$ and $\ket{d}$ of the $\Lambda$ subsystem form the internal dark state $\ket{D}=(\Omega_{4}\ket{c}-\Omega_{3}\ket{d})/ \Omega$ and the bright state $\ket{B}=(\Omega^*_{3}\ket{c}+\Omega^*_{4}\ket{d}) / \Omega$. Here, $\Omega=\sqrt{|\Omega_{3}|^2+|\Omega_{4}|^2}$ is the total Rabi frequency. 
In terms of internal dark and bright states, the total Hamiltonian transforms:
\begin{equation}
H=-\Omega^*_{p}\ket{a}\bra{b} -\beta\ket{D}\bra{b}-\alpha\ket{B}\bra{b}-\Omega\ket{B}\bra{e}+H.c.
\end{equation}
The parameters $\alpha$, and $\beta$ are defined as;
\begin{equation}
\alpha=\frac{1}{\Omega}(\Omega^*_{1}\Omega^*_{3}+\Omega^*_{2}\Omega^*_{4}).
\end{equation}
\begin{equation}
\beta=\frac{1}{\Omega}(\Omega^*_{1}\Omega^*_{4}-\Omega^*_{2}\Omega^*_{3}),
\end{equation}
The equivalent energy level diagram in transforming internal dark and bright states is shown in Fig. \ref{fig:sys} (c).
%%%%%%%%%%%%%%%%%%%%%%%%%%%%%%%%%%%%%%%%%%%%

%%%%%%%%%%%%%%%%%%%%%%%%%%%%%%%%%%%%%%%
Next, we calculate the matrix element ${\rho}_{ba}$ using the density matrix approach \cite{scully1997quantum}, which represents the optical coherence associated with the probe transition from $\ket{a}$ to $\ket{b}$. 
%%%%%%%%%
\begin{align}
\dot{\rho}_{ba} &= -(\gamma_b/2 - i\Delta_p)\rho_{ba} + i\alpha\,\rho_{ca} + i\beta\,\rho_{da} + i\Omega_p, \nonumber \\
\dot{\rho}_{ca} &= i\Delta_p \rho_{ba} + i\alpha^* \rho_{ba} + i\Omega_{ca}, \nonumber \\
\dot{\rho}_{da} &= i\Delta_p \rho_{ba} + i\beta^* \rho_{ba}, \nonumber \\
\dot{\rho}_{ea} &= -(\gamma_e/2 - i\Delta)\rho_{ea} + i\Omega_p \rho_{ba}.
\end{align}
Here $\rho_{ca}$, $\rho_{da}$, and $\rho_{ea}$ represent the ground-state coherences between $\lvert a \rangle$ and $\lvert B \rangle$, $\lvert D \rangle$, or $\lvert e \rangle$, respectively. 
The parameter $\gamma_e$ is the decay rate of the excited state $\ket{e}$ and $\gamma_b$ is the decay rate of the excited state $\ket{b}$.
As the probe field is assumed to be significantly weaker than the control fields, most of the atomic population resides in the ground state $\ket{a}$, allowing us to treat the probe field as a perturbation. We assume all the control fields are at resonance with the respective transition. Under slowly varying amplitude and steady-state conditions, the obtained density matrix element $\rho_{ba}$ becomes \cite{scully1997quantum,hamedi2017electromagnetically}
\begin{widetext}
\begin{equation}
\label{eq:rho}
\rho_{ba}=\frac{\Omega_{p} \Delta_{p}(-|\Omega|^2+i\Delta_{p}(\gamma_{e}/2-i\Delta_{p}))}{i\Delta_{p}(\gamma_{e}/2-i\Delta_{p})\zeta+i|\Omega|^2\Delta_{p}(\gamma_{b}/2-i\Delta_{p})+(\gamma_{b}/2-i\Delta_{p})(\gamma_{e}/2-i\Delta_{p})\Delta^2_{p}-|\Omega|^2|\beta|^2},
\end{equation}
\end{widetext}
%%%%%%%%%%%%%%%%%%%%%%%%%%%%%%%%%%%%
where $\zeta=|\alpha|^2+|\beta|^2$.
We define probe field detuning as $\Delta_p = \omega_p - \omega_{ab}$, where $\omega_p$ is probe field frequencies.
The optical response of the probe field is determined by the susceptibility $\chi=\eta\rho_{ba}$ of the medium with $\eta=N|\mu_{ba}|^2/\epsilon_{0}\hbar\Omega_{p}$ \cite{scully1997quantum}. Here $N$ is the number of atoms per unit volume and $\mu_{ba}$ is the dipole moment between transition $\ket{b}$ to $\ket{a}$. That indicates that $\eta$ is the parameter controllable by atomic density and considered an atomic density parameter throughout the paper. 
Therefore, the susceptibility of the five-level atomic medium and hence its permittivity $\epsilon_2$ can be modified and controlled by changing several parameters such as the $\eta$, $\Delta_p$, $\alpha$, $\beta$, and control field Rabi frequencies. 

We consider the cold atomic sample injected from the bottom nozzle into an ultra-high vacuum micro atomic cavity or glass vapor cell made of Pyrex, as shown in Fig~\ref{fig:sys}(a). 
The upper and lower layers of the pyrex glass cell have permittivity $\epsilon_1=\epsilon_3=2.25$, while inside the five-level atomic medium has permittivity $\epsilon_2$.
The inner thickness of the glass cell or cavity containing the atomic vapors is $d=0.4~\mu$m.
\footnote{Such micro vapor cells made of quartz or Pyrex are available from Akatsuki Tech Japan. A typical value of Pyrex glass refractive index is approximately 1.5, leading to $\epsilon \approx 2.25$.}
We consider that a TE and TM-polarized probe light beam is incident on the upper surface of the glass cell at an incident $\theta_{i}$. 
This monochromatic Gaussian probe beam will be reflected at the structure interface or pass through the structure. In the reflection geometry, for a TM polarized Gaussian beam reflected by the interface, the field amplitudes of two circular components of reflected light can be expressed as ~\cite{tan_enhancing_2016}:
\begin{equation}\label{a11}
\begin{aligned}
E_r^{ \pm} \propto & \frac{w_0}{w} \exp \left[-\frac{x_r^2+y_r^2}{w}\right] \\
& \times\left[r_p-\frac{2 i x_r}{k_0 w} \frac{\partial r_p}{\partial \theta_{i}} \mp \frac{2 y_r \cot \theta_{i}}{k_0 w}\left(r_p+r_s\right)\right].
\end{aligned}
\end{equation}
Here, $w=w_0\left[1+\left(2 \Lambda_{r} / k_0 w_0^2\right)^2\right]^{1 / 2}$ with beam waist $w_0$ and Rayleigh range $\Lambda_{r}=\pi w_0^2 / \lambda$. 
Here, $k_0=2 \pi / \lambda$ denotes the incident wave vector with $\lambda$ being the light wavelength.
The reflected light coordinate system is $\left(x_r, y_r, z_r\right)$, where superscript $\pm$ denotes left-hand circularly polarized (LHCP) and right circularly polarized (RHCP) photon states. 
The complex reflection coefficients for TM polarized $r_p$ and TE-polarized $r_s$ can be written as \cite{waseem2024,wan_controlling_2020}
\begin{equation}\label{a33}
r_{p, s}=\frac{r_{p, s}^{12}+r_{p, s}^{23} e^{2 i k_{2 z} d}}{1+r_{p, s}^{12} r_{p, s}^{23} e^{2 i k_{2 z} d}},
\end{equation}
where $r_{p, s}^{i j}$ is the Fresnel's reflection coefficient at the $i$-$j$ interface (here $i,j=1,2,3$).
For TM polarized
\begin{equation}\label{a44}
r_p^{i j}=\frac{k_{i z} / \varepsilon_i-k_{j z} / \varepsilon_j}{k_{i z} / \varepsilon_i+k_{j z} / \varepsilon_j},
\end{equation}
and TE polarized
\begin{equation}\label{a55}
r_s^{i j}=\frac{k_{i z}-k_{j z}}{k_{i z}+k_{j z}}.
\end{equation}
Here $k_{i z}=\sqrt{k_0^2 \varepsilon_i-k_x^2}$ represents the normal wave vector in the corresponding layer, and $k_x=\sqrt{\varepsilon_1} k_0 \sin \theta_{i}$ is the wave vector along the $x$ direction. 

It can be seen from Eq.~\eqref{a44} and Eq.~\eqref{a55} that the reflection coefficients depend on the permittivity of each layer. The upper and lower surface of the Pyrex glass cell has permittivities of 2.25.
The permittivity of the atomic medium is related to its susceptibility by the relation $\epsilon_{2}=1+\chi$. The equivalent refractive index is defined as $n=\sqrt{1+\chi}$. Here, $\chi$ represents the dielectric susceptibility of the atomic medium inside the glass cell and can be expressed as $\chi = \chi_{1} +i \chi_{2}$, such that $\chi_{1}$ represents the dispersion and $\chi_{2}$ represents the absorption of the probe field.
In our model, permittivity $\epsilon_2$ can be effectively controlled by manipulating $\chi$ of the atomic medium. This leads to a controllable photonic SHE of light.
The transverse displacements can be computed as~\cite{tan_enhancing_2016}
\begin{equation}\label{a22}
\delta^{ \pm}_{p}=\frac{\iint y_r\left|E_r^{ \pm}\left(x_r, y_r, z_r\right)\right|^2 \mathrm{~d} x_r \mathrm{~d} y_r}{\int\left|E_r^{ \pm}\left(x_r, y_r, z_r\right)\right|^2 \mathrm{~d} x_r \mathrm{~d} y_r}.
\end{equation}
It should be noted that the density matrix element $\rho_{ba}$ in Eq.~\eqref{eq:rho} is complex, and consequently, the susceptibility $\chi$ and the permittivity $\epsilon_2$ of the atomic medium are also complex. According to Eqs.\eqref{a33}--\eqref{a55}, this complexity extends to the Fresnel coefficients $r_s$ and $r_p$, giving rise to complex spin-dependent photonic SHE. 
%\Theta_{p}^{\mp}/ \Lambda_r
For horizontal polarization, the complex spin-dependent shift can be written as~\cite{mi_observation_2017, chen2021measurement}
\begin{equation}\label{eq}
\delta=\delta_{p}^{\pm} + i \delta_k^{\mp}.
\end{equation} 
%Here, $\Theta_{p}^{\mp}= \delta_k/ \Lambda_r$
Specifically, the real part of the complex shift $\delta_{p}^{\pm}$ governs the spin-dependent spatial splitting in position space. Utilizing the first-order Taylor series expansion of the Fresnel reflection coefficients, the corresponding transverse spin-displacements $\delta_p^{+}$ and $\delta_p^{-}$ can be expressed in terms of the reflective coefficients of the three-layer atomic system 
as~\cite{xiang_enhanced_2017,waseem2024}:
\begin{equation}\label{eq:shift}
\delta_p^{ \pm}=\mp \frac{k_1 w_0^2 \operatorname{Re}\left [1+\frac{r_s}{r_p} \right]  \cot \theta_{i}}{k_1^2 w_0^2+\left|\frac{\partial \ln r_p}{\partial \theta_{i}}\right|^2+\left|\left(1+\frac{r_s}{r_p}\right) \cot \theta_{i}\right|^2},
\end{equation}
with $k_{1}=\sqrt{\varepsilon_1} k_0$. Here, $\operatorname{Re}\left [1+\frac{r_s}{r_p} \right]=1+(\left|r_s\right|/ \left| r_p \right|) \cos{(\phi_s - \phi_p)}$.
Here, $\left|r_s\right|$ and $ \left| r_p \right|$ are the absolute values of the complex reflected Fresnel coefficients; $\phi_s$ and  $\phi_p$ denote their arguments.
Equation~\eqref {eq:shift} indicates that the transverse spin-dependent photonic SHE strongly depends on the reflectance intensity ratio $\left|r_s\right|/ \left| r_p \right|$. A larger ratio will result in a larger photonic SHE and vice versa.
We only presented the transverse shift of the right circularly polarized photon spin-dependent component $\delta_{p}^{+}$ because the beam shifts for the two circular components are equal in magnitude and opposite in sign.

The imaginary part of the complex shift $\delta_{k}^{\mp}$ observed at a distance $z$ from the reflection position is related to the spin-dependent splitting in momentum space, which manifests itself as an angular shift $\Theta^{\mp}=z \delta_k^{\mp}/ \Lambda_r$.
The angular spin-dependent shift can be expressed as
\begin{equation}\label{eq:angular}
\Theta_p^{ \mp}=\pm \frac{k_1 w_0^2 \operatorname{Im}\left [1+\frac{r_s}{r_p} \right]  \cot \theta_{i}}{k_1^2 w_0^2+\left|\frac{\partial \ln r_p}{\partial \theta_{i}}\right|^2+\left|\left(1+\frac{r_s}{r_p}\right) \cot \theta_{i}\right|^2} \times \frac{1}{\Lambda_r}.
\end{equation}
Here, $\operatorname{Im}\left [1+\frac{r_s}{r_p} \right]=(\left|r_s\right|/ \left| r_p \right|) \cos{(\phi_s - \phi_p)}$
Equation~\eqref{eq:shift} and Eq.~\eqref{eq:angular} reduce to the form given in Ref. \cite{mi_observation_2017} under the assumption that the last two terms in the denominator are negligible. Since $\Theta_p^{-}$ and $\Theta_p^{+}$ show the identical results with opposite symmetry. Therefore, we only consider $\Theta^{-}$ in our analysis.

\section{\label{sec:level3}Results and discussion}
This section is devoted to the numerical analysis of the photonic SHE.
To analyze photonic SHE, we select the fixed decay rates $\gamma_{e}=\gamma_{b}=\gamma$ and beam waist $w_{0}=50 \lambda$. All the other parameters are scaled with $\gamma=1$MHz.
\subsection{Spatial photonic spin Hall effect}
%%%%%%%%%%%%%%%%%%%%%%%%%%%%%%%%%%%%%%%%%%%%%%%%%%%%%%%%%%%%%%%%%%%%%%%%%%%%%%%%%%
The parameters $\alpha$, $\beta$, $\phi$ and $\eta$ are the tunable parameters. We discuss the following three cases depending on the appropriate choice of control field Rabi frequencies and their relative phase.
%%%%%%%%%%%%%%%%%%%%%%%%%%%%%%%%%%%%%%%%%%%%%%%%%%%%%%%%%%%%
%%%%%%%%%%%%%%%%%%%%%%%%%%%%%%%%%%%%%%%%%%%%%%%%%%%%%%%%%%%
\begin{figure*}[ht!]
	\centering
	\includegraphics[width=0.91\linewidth]{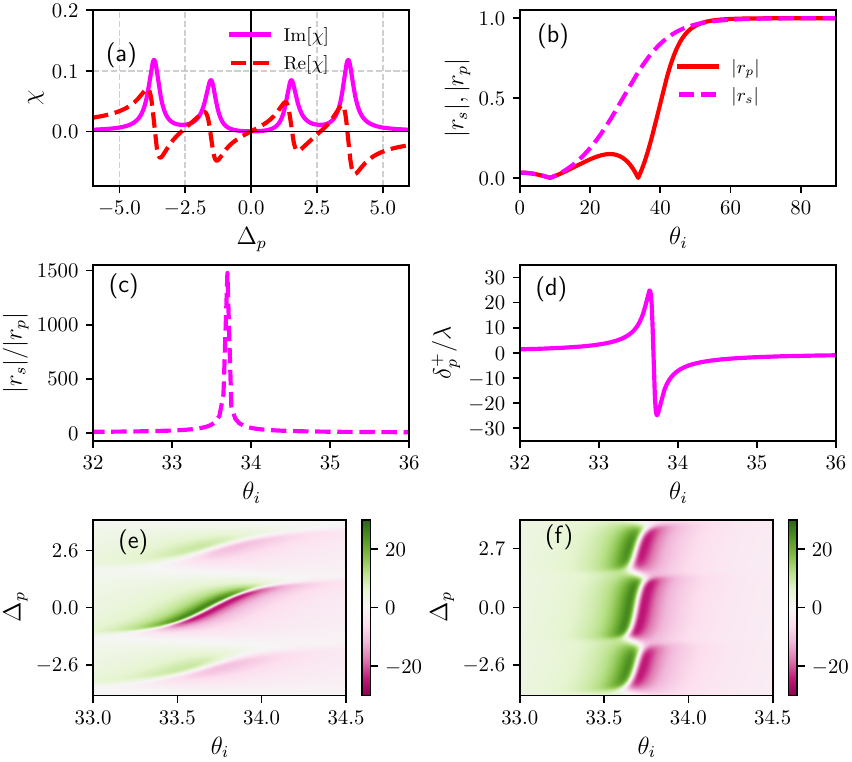}
\caption{ (a) The absorption (solid curve) and dispersion (dashed curve) characteristics of CTL atomic system as a function of probe field detuning at $|\Omega_{1}|=1.5\gamma$, $|\Omega_{2}|=3\gamma$, $|\Omega_{3}|=2.5\gamma$, $|\Omega_{4}|=0.9\gamma$, $\phi=0$, and $\eta=0.1 \gamma$.
(b) Fresnel coefficient$|r_s|$ and $|r_p|$ as a function of incident angle $\theta_{i}$ while (c) shows their respective ratio.
(d) photonic SHE changing sign from positive to negative around the angle at which the ratio $|r_s| / |r_p|$ is maximum.  
(e) Density plot of photonic SHE as a function of probe field detuning $\Delta_p$ and incident angle $\theta_i$, while (f) is a density plot when atomic density reduces one order of magnitude at $\eta=0.01 \gamma$. Note that the unit of the incident angle is degrees.} 
\label{fig:CTL}
\end{figure*}
%%%%%%%%%%%%%%%%%%%%%%%%%%%%%%%%%%%%%%%%%%%%%%%%%%%%%%%%%%%%%%%%%

\subsubsection{$\alpha \neq 0$ and $\beta \neq 0$}
%%%%%%%%%%%%%%%%%%%%%%%%%%%%%%%%%%%%%%%%%%%%%%
%%%%%%%%%%%%%%%%%%%%%%%%%%%%%%%%%%%%%%%%%%%%%%
%%%%%%%%%%%%%%%%%%%%%%%%%%%%%%%%%%%%%%%%%%%%%%
When both $\alpha$ and $\beta$ are non-zero, the five-level atomic system operates as a CTL model \cite{abbas2024goos}. To achieve this, we set the asymmetric Rabi frequencies $|\Omega_{1}|=1.5\gamma$, $|\Omega_{2}|=3\gamma$, $|\Omega_{3}|=2.5\gamma$, $|\Omega_{4}|=0.9\gamma$, and relative phase $\phi=0$, ensuring that both $\alpha$ and $\beta$ are non-zero.
We first examine the real and imaginary components of susceptibility as a function of probe field detuning, $\Delta_{p}$, in the CTL model. 
Figure~\ref{fig:CTL}(a) shows the real part of susceptibility with a dashed curve, while the imaginary part of susceptibility is shown with the solid curve. 
%The real and imaginary part of susceptibility reveals slow-light characteristics as seen by positive dispersion slope in Fig.~\ref{fig:CTL}(a). 
The imaginary part of $\chi$ also demonstrates three transparency windows in the absorption spectrum at $\Delta_p = 0$ and $\Delta_p = \pm 2.6 \gamma$.
Absorption and dispersion are zero at resonance $\Delta_{p} = 0$, similar to the phenomenon of electromagnetic-induced transparency. At detuning values of $\Delta_{p} = \pm 2.6 \gamma$, we observe another absorption window with a non-zero magnitude and zero dispersion. These results show that the refractive index, $n=\sqrt{1+\chi}$, can be tuned by adjusting the probe field detuning, allowing control over the photonic SHE.

Next, we consider the situation when the probe field is incident at angle $\theta_{i}$ as depicted in Fig.~\ref{fig:sys}(a) to make clear the relationship between the ratio $|r_s|/|r_p|$ and the photonic SHE.
To observe the angle around which ratio $|r_s|/|r_p|$ becomes large, we first show the Fresnel coefficient$|r_s|$ and $|r_p|$ as a function of incident angle $\theta_{i}$ at $\Delta_{p}=0$, and $\eta=0.1 \gamma$, as shown in Fig.~\ref{fig:CTL}(b).
The figure clearly shows that $|r_s|$ is monotonic increasing function of $\theta_i$ and becomes saturated after $\theta_i \approx 50^\circ$.
At an incident angle of $\theta_{i}=33.7^\circ$, the Fresnel coefficient $|r_p|$ approaches zero while $|r_s|$ maintains a finite value, and this angle is referred to as the Brewster angle $\theta_B$. 
This strong reduction of the $|r_p|$ around the Brewster angle $\theta_B$ is due to the formation of an evanescent wave which propagates along the interface between the atomic medium and cavity~\cite{salasnich_enhancement_2012}.
At this Brewster angle, the ratio $|r_s|/|r_p|$ significantly increases as seen in Fig.~\ref{fig:CTL}(c).

Figure~\ref{fig:CTL}(d) further shows the prominent photonic SHE around Brewster's angle $\theta_{B}=33.7^\circ$.
The photonic SHE switches sign from a positive peak value of 25$\lambda$ to a negative peak value -25$\lambda$ around $\theta_{i}=33.7^\circ$. 
The peak value is exactly half of the beam waist of the incident beam, which is the maximum upper limit of photonic SHE.
The transverse photonic SHE is positive for $\theta_{i} < 33.7^\circ$, and negative for $\theta_{i} > 33.7^\circ$. 
One possible reason for this change in sign of photonic SHE is that the horizontal component of the probe’s electric field switches its phase while the vertical component remains unchanged. Therefore, the phase difference $\phi_s  - \phi_p$ associated with $|r_s|$ and $|r_p|$ experiences a $\pi$ phase variation, and the spin accumulation would reverse its directions accordingly~\cite{waseem2024}.

Both the reduction of $|r_P|$, the ratio $|r_s|/|r_p|$, and the position of Brewster angle $\theta_B$ depend on the parameters of the atomic system, which influences the photonic SHE. To study the effects of probe field detuning on photonic SHE, figure~\ref{fig:CTL}(e) shows a density plot of the photonic SHE as a function of detuning and incident angle at fixed $\eta=0.1 \gamma$. A maximum photonic SHE of 25$\lambda$ appears at $\Delta_p = 0$, with an additional, lower-magnitude (around $\leq 10 \lambda$) photonic SHE at $\Delta_p = \pm 2.6 \gamma$ due to non-zero absorption at these detuning values.
Another noteworthy effect is that the angle at which $|r_p|$ approaches zero shifts by approximately $\pm1^\circ$ from $33.7^\circ$. 
Figure~\ref{fig:CTL}(f) shows a density plot of photonic SHE when the atomic density is reduced by an order of magnitude, setting $\eta=0.01 \gamma$. With this lower density, photonic SHE enhances compared to higher density and reaches 20$\lambda$ at $\Delta_p = \pm 2.6 \gamma$ while remaining the same value of 25$\lambda$ at $\Delta_p = 0$.
Fig.~\ref{fig:CTL} (e) and (f), it is evident that at $\Delta_p = 0$, photonic SHE is independent of density parameter $\eta$ and stays at a constant value of $\pm 25 \lambda$. 
At $\Delta_{p} = \pm 2.6 \gamma$, reducing the atomic density minimizes probe field absorption, further enhancing the photonic SHE. 
Additionally, the range of incident angles over which the photonic SHE changes sign from positive to negative becomes much narrower for $\eta=0.01 \gamma$ compared with $\eta=0.1 \gamma$.
Furthermore, photonic SHE at probe field resonance is independent of control field strength as long as the EIT condition is satisfied due to zero absorption and dispersion.

%%%%%%%%%%%%%%%%%%%%%%%%%%%%%%%%%%%%%%%%%%%%%%%
%%%%%%%%%%%%%%%%%%%%%%%%%%%%%%%%%%%%%%%%%%%%%%%%%%%%%%
%%%%%%%%%%%%%%%%%%%%%%%%%%%%%%%%%%%%%%%%%%%%%%%%%%%%%%%%%%%%%%%%%%%%%%%%

%%%%%%%%%%%%%%%%%%%%%%%%%%%%%%%%%%%%%%%%%
%%%%%%%%%%%%%%%%%%%%%%%%%%%%%%%%%%%%%%%%%%%%%%%%%%%%%%%%%%%%%%%%%%%
\subsubsection{$\alpha=0$ and $\beta \neq 0$}
\begin{figure*}[htbp]
	\centering
\includegraphics[width=0.990\linewidth]{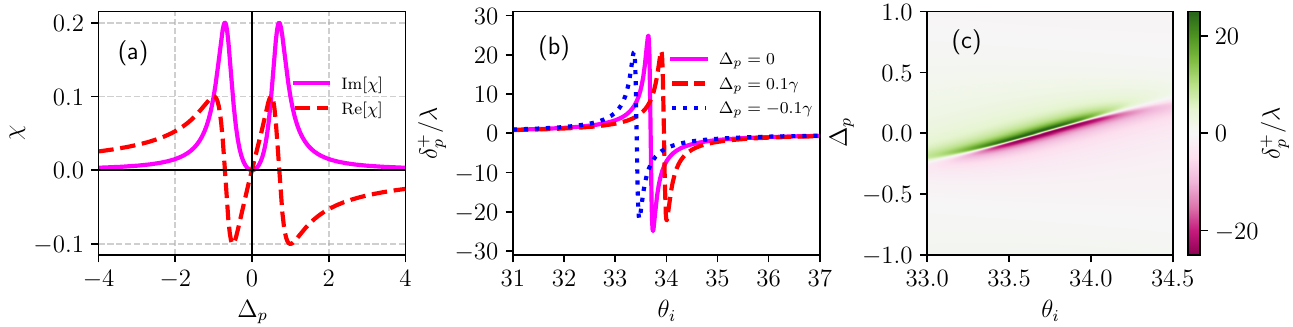}
\caption{(a) The absorption (solid curve) and dispersion (dashed curve) characteristics as a function of probe field detuning at $|\Omega_1 |= |\Omega_2|=0.5 \gamma$, $|\Omega_3 |= |\Omega_4|=0.7 \gamma$, $\phi=\pi$ and $\eta=0.1 \gamma$. These symmetric Rabi frequency values create the situation $\alpha=0$ and $\beta \neq 0$. As a result, the atomic system becomes an effective $\Lambda$ system showing the EIT-like characteristics of the probe field.  (b) photonic SHE as a function of probe field incident angle at three different values of $\Delta_p$. (c) Two-dimensional density plot of photonic SHE as a function of probe field detuning and incident angle at $\eta=0.1 \gamma$. Note that the unit of the incident angle is degrees.} 
\label{fig:eit}
\end{figure*}

%%%%%%%%%%%%%%%%%%%%%%%%%%%%%%%%%%%%%%%%%%%
%%%%%%%%%%%%%%%%%%%%%%%%%%%%%%%%%%%%%%%%%%%
When $\alpha$ is zero and $\beta$ is non-zero, the five-level atomic system transforms to a conventional $\Lambda$-type configuration. This change is due to the decoupling of levels $\ket{B}$ and $\ket{e}$, as shown in Fig.~1(c). To achieve $\alpha=0$ and $\beta \neq 0$, we consider a symmetric setup with $|\Omega_1|=|\Omega_2|=0.5 \gamma$ and $|\Omega_3|=|\Omega_4|=0.7 \gamma$, and set the relative phase to $\phi=\pi$.
Figure~\ref{fig:eit}(a) illustrates the susceptibility as a function of probe field detuning.  
This spectral response reflects the standard characteristics of electromagnetic-induced transparency (EIT) and slow light, where the intracavity medium displays EIT with normal dispersion.
The plots of $|r_s|$, $|r_p|$, and $|r_s | / |r_p |$ show a similar trend to the CTL case and are not presented for simplicity.
The solid curve in Fig.~\ref{fig:eit}(b) shows that the photonic SHE behavior at $\Delta_{p} = 0$ is identical to that of the CTL atomic configuration. Because, in both cases, the corresponding refractive index is unity with total $\chi = 0$. 
However, in the nearby region around resonance at $\Delta_p = 0$, the refractive index varies with detuning due to changes in susceptibility. The dashed and dotted curves illustrate the photonic SHE at detunings of $\Delta_p = 0.1 \gamma$ and $\Delta_p = -0.1 \gamma$, respectively, showing the shift in the angle at which the photonic SHE changes sign from positive to negative.
To further analyze these findings, Fig.~\ref{fig:eit}(c) presents a density plot of the photonic SHE versus incident angle $\theta_{i}$ and probe field detuning $\Delta_p$. 
In contrast to the CTL case, maximum value of photonic SHE is limited to the resonance region around $\Delta_p = 0$ due to a single transparency window. Thus, photonic SHE is maximum at zero probe field detuning, which is similar to the CTL case. 
This photonic SHE is independent of the control field strength and atomic density as long as the EIT condition holds. In other words, the complex CTL-type system and the simpler $\Lambda$-type system exhibit similar photonic SHE characteristics at probe field resonance.

%%%%%%%%%%%%%%%%%%%%%%%%%%%%%%%%%%%%%%%%%%%%%%
%%%%%%%%%%%%%%%%%%%%%%%%%%%%%%%%%%%%%%%%%

\subsubsection{$\alpha \neq 0$ and $\beta = 0$}

\begin{figure*}[ht!]
	\centering
\includegraphics[width=1\linewidth]{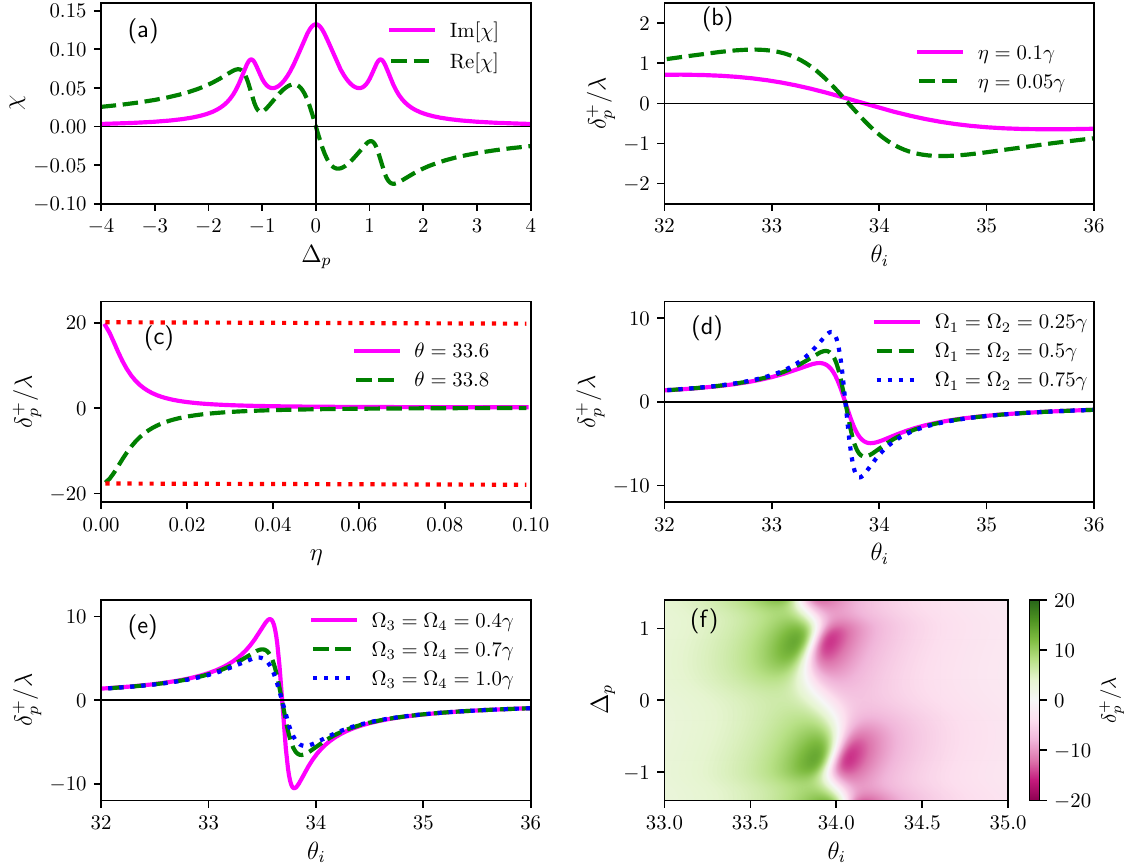}
\caption{ (a) The absorption (solid curve) and dispersion (dashed curve) characteristics as a function of probe field detuning at $|\Omega_1 |= | \Omega_2|=0.5 \gamma$, $|\Omega_3| = |\Omega_4|=0.7 \gamma$, $\phi=0$, and $\eta=0.1 \gamma$. These symmetric control field values create the situation $\alpha \neq 0$ and $\beta = 0$, and the atomic system becomes effectively an $N$ type.
(b) photonic SHE as a function of the incident angle at $\Delta_{p} =0$ at two different values of $\eta = 0.1 \gamma$ and $\eta=0.05 \gamma$. 
(c) photonic SHE dependence on density parameter $\eta$ at $\Delta_p =0$ and at two different values of incident angles. The horizontal dotted line shows no dependence on $\eta$ for CTL and EIT cases. 
(d) photonic SHE as a function of the incident angle at $\Delta_p = 0$ at three different values of $|\Omega_{1}|=|\Omega_{2}|$ whereas (e) is at three different values of $|\Omega_{3}|=|\Omega_{4}|$.
(f) Two-dimensional plot of photonic SHE as a function of incident angle and probe field detuning. Note that the unit of the incident angle is degrees.} 
\label{fig:N}
\end{figure*}

%%%%%%%%%%%%%%%%%%%%%%%%%%%%%%%%%%%%%%%%%%%%%%%%%%%%%%
When $\alpha$ is non-zero and $\beta$ is zero, the five-level atomic system behaves as a conventional $N$-type atomic system due to the decoupling of level $\ket{D}$, as shown in Fig.~\ref{fig:sys}(c). To achieve this condition ($\alpha \neq 0$ and $\beta = 0$), we set up a symmetric configuration with $|\Omega_1|=|\Omega_2|=0.5 \gamma$ and $|\Omega_3|=|\Omega_4|=0.7 \gamma$, while keeping a relative phase of $\phi=0$. 
Figure~\ref{fig:N}(a) displays the susceptibility as a function of the probe field detuning. 
%The negative slope in the dispersion indicates superluminal behavior. 
Unlike the CTL and EIT cases, the dispersion is zero while absorption remains non-zero at resonance $\Delta_p = 0$. 
The plots of $|r_s|$, $|r_p|$, and the ratio $|r_s| / |r_p|$ follow trends similar to the CTL and EIT cases, so they are omitted here for simplicity. However, it is worth noting that the peak value of the ratio $|r_s| / |r_p|$ is nearly two orders of magnitude smaller and exhibits a broader full-width at half maximum.

In Fig.~\ref{fig:N}(b), photonic SHE is shown as a function of the incident angle at $\Delta_p = 0$ for two atomic density parameters $\eta=0.1 \gamma$ (solid curve) and $\eta = 0.05 \gamma$ (dashed curve). Compared to the CTL and EIT cases, the magnitude of photonic SHE is smaller, and the range of angles where the sign changes is less steep. For the lower density, $\eta = 0.05 \gamma$, the photonic SHE is larger than at $\eta = 0.1 \gamma$. 
To further analyze this effect, Fig.~\ref{fig:N}(c) shows photonic SHE as a function of atomic density $\eta$ at two fixed incident angles: $\theta_{i}=33.6^\circ$ (corresponding to the maximum positive shift) and $\theta_{i}=33.7^\circ$ (corresponding to the minimum shift). 
As the atomic density parameter $\eta$ increases, absorption also increases. This leads to an increased value of the imaginary part of the susceptibility. As a result, the $|r_s| / |r_p|$ ratio is reduced, leading to a decrease in photonic SHE. Therefore, a lower atomic density is more suitable for enhancing photonic SHE in an $N$-type atomic system at $\Delta_p = 0$. This contrasts with the EIT and CTL cases at resonance, where atomic density does not affect photonic SHE, as indicated by the horizontal lines in Fig.~\ref{fig:N}(c).
Unless stated otherwise, we fix $\eta = 0.01 \gamma$ in the rest of this section.

Next, we explore how varying the Rabi frequencies of the control fields affects photonic SHE, ensuring that $\alpha$ and $\Omega$ remain nearly comparable so that the system stays in an $N$-type configuration. 
First, we fix $|\Omega_{3}|=|\Omega_{4}|=0.7 \gamma$ and vary $|\Omega_{1}|=|\Omega_{2}|$. The solid, dotted, and dashed curves in Fig.~\ref{fig:N}(d) represent the photonic SHE as a function of the incident angle for $|\Omega_{1}|=|\Omega_{2}|=0.25 \gamma$, $|\Omega_{1}|=|\Omega_{2}|=0.5 \gamma$, and $|\Omega_{1}|=|\Omega_{2}|=0.75 \gamma$, respectively. The results show that photonic SHE increases with the increase of the control field strength $|\Omega_1| = |\Omega_2|$.
We then set $|\Omega_{1}|=|\Omega_{2}|=0.5 \gamma$ and vary $|\Omega_{3}|=|\Omega_{4}|$. The solid, dotted, and dashed curves in Fig.~\ref{fig:N}(e) display the behavior of photonic SHE as a function of the incident angle for $|\Omega_{3}|=|\Omega_{4}|=0.4 \gamma$, $|\Omega_{3}|=|\Omega_{4}|=0.7 \gamma$, and $|\Omega_{3}|=|\Omega_{4}|=1.0 \gamma$, respectively. In this case, photonic SHE increases as the control field $|\Omega_3|=|\Omega_4|$ decreases. These results are different than the CTL and EIT cases, where control field amplitude does not influence photonic SHE at resonance. 
%\begin{widetext}
%\end{widetext}
%%%%%%%%%%%%%%%%%%%%%%%%%%%%%%%%
Finally, the two-dimensional density plot in Fig.~\ref{fig:N}(f) shows photonic SHE as a function of probe field detuning and incident angle. photonic SHE reaches higher values at $\Delta_p \approx \pm 1 \gamma$ than at $\Delta_p = 0$. This increase is due to two lower absorption dips at $\Delta_p \approx \pm 1 \gamma$ than at $\Delta_p = 0$, as seen in the susceptibility curve in Fig.~\ref{fig:N}(a).
This behavior contrasts with the CTL case, where photonic SHE smoothly decreases as it moves away from the resonance point $\Delta_p = 0$.   
Overall, the magnitude of photonic SHE is lower at $\Delta_p = 0$ and higher at $\Delta_p \neq 0$, while in the CTL case, it behaves oppositely, as evidenced by comparing these results with Fig.~\ref{fig:CTL}(e).

\subsection{Angular photonic spin Hall effect}
\begin{figure}[htbp]
	\centering
\includegraphics[width=0.90\linewidth]{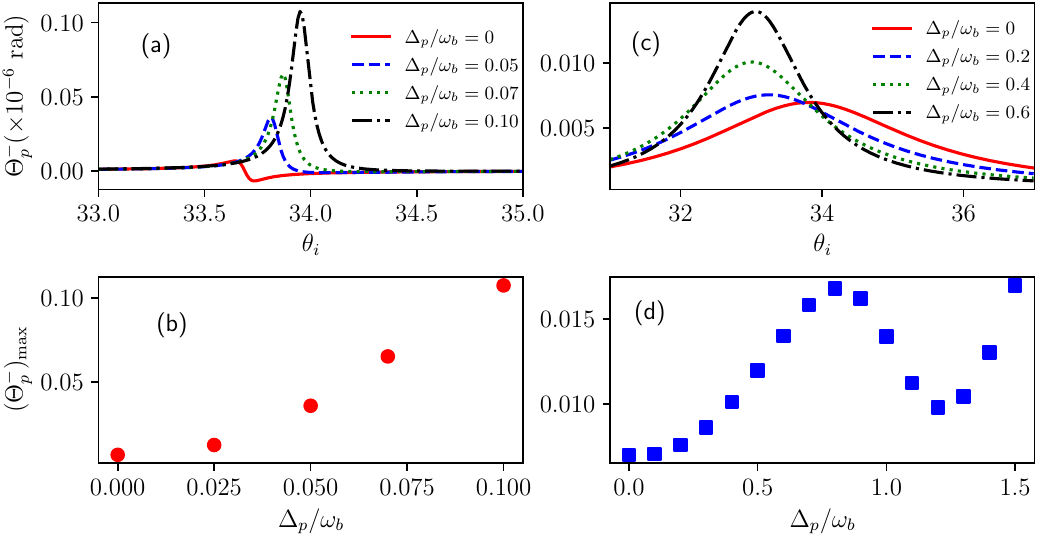}
\caption{(a) Angular photonic SHE versus incident angle for $\Lambda$-type EIT system at three different values of $\Delta_p$.  (b) Maximum value of Angular photonic SHE versus probe field detuning for $\Lambda$-type. (c) Angular photonic SHE as a function of incident angle under $N$-type configuration at three different values of $\Delta_p$. (c) Maximum value of angular photonic SHE versus prob field detuning for $N$-type system. Note that the unit of the incident angle is degrees.} 
\label{fig:angular}
\end{figure}

%10%%%%%%%%%%%%%%%%%%%%%%%%%%%%%%%%%%%%%%%%%%%%%
%%%%%%%%%%%%%%%%%%%%%%%%%%%%%%%%%%%%%%%%%%%%%%
Next, we explore the angular spin-dependent shift in our system. We first consider the $\Lambda$-type EIT system ($\alpha=0$ and $ \beta \neq 0$) and explore the angular shift behavior in the vicinity of zero probe field detuning. 
Figure~\ref{fig:angular}(a) shows the angular shift $\Theta^{-}$ as a function of incident angle $\theta_i$ at three different probe field detuning. 
Under exact EIT conditions ($\Delta_p = 0$), the angular shift exhibits behavior analogous to the spatial shift, as demonstrated by the red solid curve in Fig.~\ref {fig:angular}(a).

For non-zero detuning ($\Delta_p \neq 0$), the angular shift maintains strictly positive values, in contrast to the spatial shift that exhibits sign reversal.
The maximum value of the angular shift increases with the increase of probe field detuning, as depicted in Fig.~\ref{fig:angular}(b).
Since changing the probe field detuning changes the susceptibility $\chi$ or refractive index of the atomic medium. The characteristic trend observed in Fig.~\ref{fig:angular}(b) serves as a direct signature for probing the refractive index of the atomic medium. It may be worth noting that the angular shift is related to the weak value amplification factor in the real experiment \cite{chen2021measurement}.
The CTL system ($\alpha \neq 0$ and $ \beta \neq 0$) near zero probe field detuning shows a similar trend of angular shift (data not shown for simplicity).
%%%%%%%%%%%%%%%%%%%%%%
%%%%%%%%%%%%%%%%%%%%%%%%%%%%%%%%%%%%%%%%%%%%%%%%%%%%%%%%%%%%
%%%%%%%%%%%%%%%%%%%%%%%%%%%%%%%%%%%%%%%%
Additionally, we also consider the angular shift for the $N$-type system for which ($\alpha \neq 0$ and $ \beta \neq 0$). Figure~\ref{fig:angular}(c) shows the angular shift versus incident angle. Unlike the behavior of the spatial shift [Fig.~\ref{fig:N}(f)], the angular photonic SHE exhibits a strictly positive shift across all probe detunings  (see Fig.~\ref{fig:angular}(c)). Figure~\ref{fig:angular}(d) presents the maximum value of the angular spin-dependent shift as a function of probe field detuning, which indicates the oscillatory behaviour due to the multiple transparency windows. 
The angular photonic SHE for the $\Lambda$-type EIT system is two orders of magnitude higher than the $N$-type system.
We choose $\eta=0.1$ for all the results shown in Fig.~\ref{fig:angular}.

\section{Conclusion}
It is worth noting that the $\Lambda$ and $N$-type atomic configurations considered here are limiting cases derived from the CTL configuration. The results presented for $\Lambda$ and $N$-type atomic systems are also applicable to natural $\Lambda$ ( i.e, $\Omega_2 = \Omega_3 = \Omega_4=0$) and $N$-type (i.e,$\Omega_3 = \Omega_4=0$) atomic structures by simply turning off the unnecessary control fields making our finding quite general for $\Lambda$ and $N$-type atomic system.
Our proposed results may be realized in the experiment by incorporating the standard weak measurement protocol~\cite{luo_enhanced_2011} on hyperfine energy levels of Alkali metals such as \textsuperscript{87}Rb.
For \textsuperscript{87}Rb atoms, the proposed implementation involves a combined tripod and $\Lambda$ configuration in the hyperfine structure. Considering hyperfine structure of \textsuperscript{87}Rb, CTL system includes two excited states, $\lvert b \rangle = \lvert 5P_{3/2}, F = 1, m_F = 0 \rangle$, and $\lvert e \rangle = \lvert 5P_{3/2}, F = 0, m_F = 0 \rangle$, as well as three lower states, $\lvert a \rangle = \lvert 5S_{1/2}, F = 2, m_F = 1 \rangle$, $\lvert c \rangle = \lvert 5S_{1/2}, F = 1, m_F = 1 \rangle$, and $\lvert d \rangle = \lvert 5S_{1/2}, F = 1, m_F = -1 \rangle$.
The ground state driven by $\Omega_p$ can be isolated from the control fields $\Omega_1$, $\Omega_2$, $\Omega_3$, and $\Omega_4$ by choosing the proper combination of the polarization configuration for probe and control fields. 
Another possible way to implement the CTL system in \textsuperscript{87}Rb hyperfine energy level is the use of a magnetic field to break the degeneracy. Laser frequencies to derive the hyperfine transition can be potentially generated using the acoustic optical modulators and electro-optical modulators. However, compared to traditional $\Lambda$ and $N$-type atomic configurations, tuning the control field resonant to the respective transition requires careful adjustment and may be challenging in real experiments. We assume a weak probe field, which allows us to linearize the density matrix equations and consider only the first-order terms in $\Omega_p$.
Under this weak probe field approximation, the total decay rates $\gamma_b$ and $\gamma_e$ are relevant, representing the sum of the branching ratios from the excited states to all lower states. 
This simplifies our analysis by focusing on the primary effects of probe field characteristics on the photonic SHE, without considering specific decay pathways. While $\gamma_b=\gamma_e=\gamma$ is assumed for \textsuperscript{87}Rb, including individual branching ratios would provide a more detailed description. 
Therefore, the experimental realization of the CTL system's photonic SHE may be challenging compared to the conventional $\Lambda$ and $N$-type systems, whose detailed description is beyond the scope of this article.
Furthermore, it is possible to avoid the Doppler broadening effects by injecting cold atomic gases or employing the Doppler-free technique~\cite{doplerfree}.

In conclusion, we studied the photonic SHE of a probe field induced by four coherent control fields. These control fields form a combined tripod and $\Lambda$ configuration. By appropriately selecting the amplitudes and phases of the control fields, a five-level atomic system can be transformed into $\Lambda$ and $N$-type atomic configurations. We demonstrated that the photonic SHE can be tuned via probe field detuning, the magnitudes of the control field Rabi frequencies, and atomic density.
At resonance, the CTL and $\Lambda$ configurations exhibited similar behavior, showing no dependence on atomic density or control field strengths, and maximum photonic SHE up to half of the incident beam waist is achieved. However, unlike the $\Lambda$ system, the CTL system displayed maximum photonic SHE at multiple probe field detunings. For the $N$-type system, the photonic SHE across all probe field detunings depends on atomic density and control field strength, providing greater tunability with a wider range of adjustable parameters.

 \section{acknowledgments}
M. Shah acknowledges financial support from the postdoctoral research grant YS304023905. 
  
%%%%%%%%%%%%%%%%%%%%%%%%%%%%%%%%%%%%%%%%%%%%%%%%%%%%%%%%%%%%%%
 \section*{Conflict of interest}
 The authors declare that they have no conflict of interest.
%%%%%%%%%%%%%%%%%%%%%%%%%%%%%%%%%%%%%%%%%%%%%%%%%%%%%%%%%%%%%%%

 \section*{Data availability}
Data underlying the results presented in this paper may be obtained from the authors upon reasonable request and are also reproducible from theoretical models.

%%%%%%%%%%%%%%%%%%%%%%% References %%%%%%%%%%%%%%%%%%%%%%%%%

%%%%%%%%%% If using BibTeX:

\bibliography{sample.bib}

% \end{thebibliography}

\end{document}